\documentclass[
reprint,
superscriptaddress,
showpacs,
preprintnumbers,
amsmath,
amssymb,
amsart,
aps, prl
floatfix,
twocolumn, 
longbibliography
] {revtex4-2}

\usepackage{graphicx}  
\usepackage{dcolumn} 
\usepackage{bm}
\usepackage{float}
\usepackage{siunitx}
\usepackage{xcolor}
\usepackage{comment}
\usepackage{bm}
\usepackage{amsthm}
\usepackage{wasysym}
\usepackage{soul}
\usepackage{upgreek}

\begin{document}

\title{Dispersive Properties of Plasma Diffraction Gratings:\\ Towards Plasma-Based Laser Pulse Compression}

\newcommand{\equalcontrib}{These authors contributed equally to this work.}

\author{V. M. Perez-Ramirez}
\email{vpr11@stanford.edu}
\thanks{\equalcontrib}
\affiliation{Stanford University, Stanford, CA 94305}
 
\author{M. M. Wang}
\email{mw2857@princeton.edu}
\thanks{\equalcontrib}
\affiliation{Princeton University, Princeton, NJ 08544}

 \author{K. Ou}
\affiliation{Stanford University, Stanford, CA 94305}

\author{S. Cao}
\affiliation{Stanford University, Stanford, CA 94305}

\author{D. Singh}
\affiliation{Stanford University, Stanford, CA 94305}

\author{N. M. Fasano}
\affiliation{Princeton University, Princeton, NJ 08544}

\author{V. Dewan}
\affiliation{Princeton University, Princeton, NJ 08544}

\author{A. M. Giakas}
\affiliation{Princeton University, Princeton, NJ 08544}

\author{A. Das}
\affiliation{Princeton University, Princeton, NJ 08544}

\author{I. Tigges-Green}
\affiliation{Princeton University, Princeton, NJ 08544}

\author{P. Michel}
\affiliation{Lawrence Livermore National Laboratory, Livermore, CA, 94551}

\author{J. M. Mikhailova}
\email{j.mikhailova@princeton.edu}
\affiliation{Princeton University, Princeton, NJ 08544}
 
\author{M. R. Edwards}
\email{mredwards@stanford.edu}
\affiliation{Stanford University, Stanford, CA 94305}

\date{\today}

\begin{abstract}
The standard architecture for a high-peak-power femtosecond laser is chirped pulse amplification using diffraction gratings for compression; the damage threshold of the compression gratings limits current lasers to multi-petawatt peak power. Plasma gratings have orders-of-magnitude higher damage tolerance than conventional optics, so plasma gratings with sufficiently high optical quality could allow the construction of ultra-high-power femtosecond lasers. Here, we present experimental measurements of the angular dispersion, angular bandwidth, and diffraction angles of ionization-based plasma transmission gratings and show that both the dispersive and the diffractive properties of these gratings are in close agreement with optical theory and simulations. Gratings with a period of $10.2 \ \mathrm{\upmu m}$ are found to have an angular dispersion of approximately $0.005^{\circ}/\mathrm{nm}$. The dispersion and bandwidth of these gratings suggest plausible designs for a plasma-grating-based compressor and indicate a pathway to compact lasers with petawatt to exawatt peak power. 
\end{abstract}


\maketitle

Chirped-pulse amplification (CPA) \cite{Strickland1985compression} dramatically increased the peak power that could be delivered by femtosecond-pulsed lasers; modern systems can produce up to 10 PW \cite{Lureau2020ten,Jourdain2020the}. The resulting increase in laser intensity has enabled laboratory study of astrophysical phenomena \cite{Zakharov2003collisionless,Fiuza2012weibel,Chen2015scaling,Takabe2021recent}, compact x-ray and extreme ultraviolet sources \cite{Gibbon1996harmonic,Dromey2013coherent,Brambrink2016short,Edwards2020x}, laser-driven particle accelerators \cite{Tajima1979laser,Geddes2004high,Esarey2009physics,Leemans2006gev,Albert2016applications,Gong2024laser}, and tests of strong-field quantum electrodynamics \cite{DiPiazza2012extremely,Weber2017p3,Meuren2020seminal,Mirzaie2024all}.
However, the standard CPA architecture is difficult to extend to hundred-petawatt or exawatt powers because the limited damage threshold of conventional gratings ($10^{12}\ $to$ \ 10^{13} \ \mathrm{W/cm^2}$ for femtosecond pulses \cite{Poole2013femtosecond,Ristau2009laser}) means that such high powers require impractically large and expensive components. Optical damage is especially relevant for the final compression gratings of CPA systems, which must withstand the full system power and are among the most expensive components of high-power lasers. The size and expense of the compression gratings required for multi-petawatt lasers is a significant obstacle to the consistent operation of current systems and further increases in peak power and repetition rate. 

The extreme robustness of plasma to high-intensity fields offers a solution. Plasma optics can tolerate intensities several orders-of-magnitude higher than solid-state equivalents; therefore, a plasma-optic-based laser could deliver higher power from a smaller footprint.
Plasma optics have already found applications in high-power laser science: plasma mirrors \cite{Thaury2007plasma,Dromey2004plasma,Scott2015optimization} are used to increase temporal contrast of intense pulses \cite{Kapteyn1991prepulse,Backus1993prepulse,Mikhailova2011ultra,Roedel2011high,Edwards2020multi,Kim2011spatio,Levy2007double,Inoue2016single} and to provide efficient nonlinear frequency conversion to second, third \cite{Edwards2020multi}, and higher-order harmonics \cite{Mikhailova2005generation, Mikhailova2012isolated, Heissler2015multi, Dromey2006high, Heissler2012few}, cross-beam energy transfer improves the symmetry of inertial fusion implosions \cite{Michel2009tuning, Glenzer2010symmetric,Moody2012multistep}, and plasma waveguides enhance electron acceleration \cite{Sheng2005plasma,Lemos2018guiding,Miao2020optical, Oubrerie2022controlled,Shrock2022meter}.
Additionally, computational studies and proof-of-principle experiments have been conducted on a variety of plasma amplifiers \cite{Malkin1999fast,Andreev2006short,Lancia2010experimental,Vieux2011chirped,Marques2019joule}, gratings \cite{Sheng2003plasma,Suntsov2009femtosecond,Lehmann2016transient,Edwards2022plasma}, lenses \cite{Ren2001compressing,Palastro2015plasma,Lehmann2019plasma,Edwards2022holographic}, and polarizers \cite{Turnbull2016high,Turnbull2017refractive,Michel2020polarization}. However, a robust and efficient plasma-based method for compressing high-power laser pulses has not yet been experimentally demonstrated. 

Recent theoretical and computational work has suggested that plasma formed into gratings could be used for temporal compression \cite{Edwards2022plasma,Lehmann2024plasma}. In particular, the plasma-grating compressor described in Ref.~\cite{Edwards2022plasma} relies on the angular dispersion of plasma transmission gratings to produce a difference of vacuum-propagation distance for different wavelengths in a four-grating approach analogous to conventional solid-state grating compressors \cite{Strickland1985compression,Treacy1969optical}.
This approach is relatively robust to plasma properties since dispersion from a transmission grating depends only on the grating period and is insensitive to density and temperature. Despite the expected robustness, validating that plasma gratings do exhibit the dispersive properties predicted by theory is necessary for building a compressor. 
Here, we present an experimental characterization of the dispersive and angular bandwidth properties of ionization gratings, comparing their behavior to simulations and theory as a step towards a compact plasma-based laser compressor. 

\begin{figure*}[t]
    \centering
    \includegraphics[width=1\linewidth]{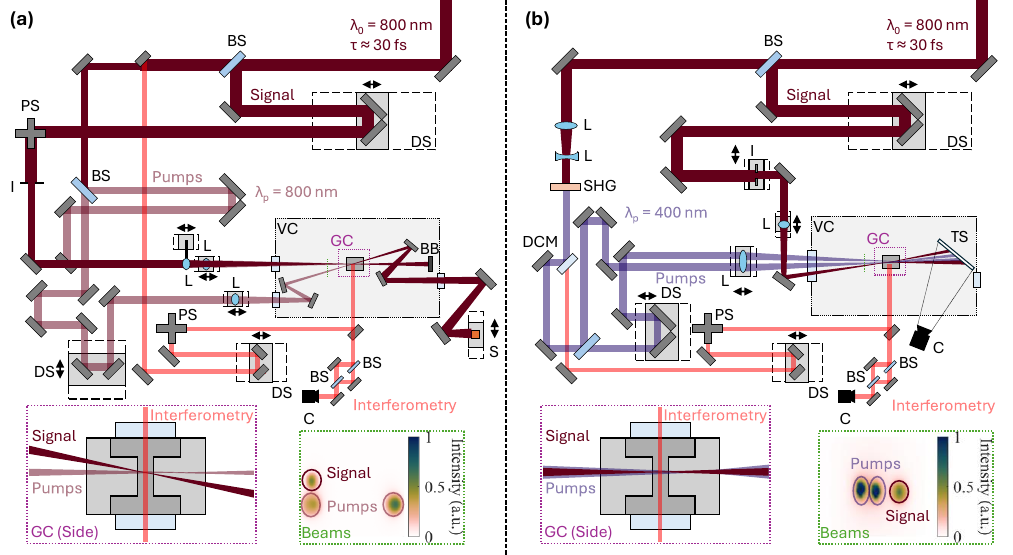}
    \caption{Schematics of experiments used to measure (a) angular dispersion and (b) angular bandwidth. Side views of the gas cell show how the signal beam propagates above the pump beam plane in configuration (a) and within the pump beam plane in configuration (b). A transverse view of the beams before the gas cell is also shown for each configuration. Component abbreviations: (BS) beam splitter, (PS) periscope, (DS) delay stage, (L) lens, (I) iris, (VC) vacuum chamber, (GC) gas cell, (BB) beam block, (DCM) dichroic mirror, (S) spectrometer, (TF) Teflon screen, and (C) camera.}
    \label{fig:1}
\end{figure*}

To create a plasma grating, we crossed two moderately intense femtosecond pump pulses (wavelength $\lambda_p$) at a full-angle $2\theta_p$ in a gas cell, producing an interference pattern with period $\Lambda = \lambda_p/(2\sin[\theta_p])$. If the intensity in the constructive interference fringes is sufficient to ionize the gas, the pumps will create a spatially varying ionization grating, where neutral gas and plasma form alternating layers \cite{Suntsov2009femtosecond,Yang2010femtosecond,Liu2010energy,Shi2011generation,Durand2011experimental,Liu2011two,Durand2012dynamics,Edwards2023control,Edwards2024greater,Wang2026experimental}. The index of refraction ($n$) of a plasma depends on the electron density as $n_{\textup{plasma}} = \sqrt{1-n_e/n_c}$ (where $n_e$ is the electron number density, $n_c = \epsilon_0m_e\omega^2/e^2$ is the plasma critical density, $\epsilon_0$ is the vacuum permittivity, $m_e$ is the mass of an electron, $\omega$ is the angular frequency of the electromagnetic wave, and $e$ is the fundamental charge) and differs from that of neutral gas at the same density ($n_{\textup{gas}}>1$), so these gas-plasma layers produce an index of refraction modulation along one dimension, $n(x) = n_0 + \delta n (x)$, where the index has been divided into average ($n_0$) and fluctuating ($\delta n$) components. For volumetric gratings in the Bragg regime \cite{Moharam1978criterion}, only the fundamental Fourier component of the index variation ($n_1$), corresponding to a sinusoidal modulation with period $\Lambda$, governs grating performance \cite{Yeh1993introduction}. For such a grating, a monochromatic beam with wavelength $\lambda$ incident at the Bragg angle, $\theta_B = \arcsin(\lambda/\left[2\Lambda\right])$, will diffract into the first order with efficiency:
\begin{equation}
\eta = \sin^2\left( \kappa L \right),
\label{eq:0}
\end{equation}
where $\kappa = \pi n_1/(\lambda\cos\left[\theta_B\right])$ is the coupling parameter and $L$ is the grating length. With appropriate choice of $n_1$ and $L$, $\eta$ can equal 1, representing total diffraction to the first order. Reported experimental work has shown average efficiencies above 35\% \cite{Edwards2023control,Wang2026experimental}.

Transmission gratings will exhibit angular dispersion for polychromatic light, as can be derived from the grating equation:
\begin{equation}
    \sin \left( \theta_i\right) + \sin \left( \theta_d\right) = \frac{m\lambda}{\Lambda},
    \label{eq:1}
\end{equation}
where $\theta_i$ is the incident angle, $\theta_d$ is the diffracted angle, m is the diffraction order, and $\lambda$ is the wavelength of light incident on the grating. For first-order diffraction ($m=1$) and small angles ($\sin\left[\theta\right] \approx \theta$), the dispersion depends primarily on the grating period \cite{Yeh1993introduction}:
\begin{equation}
    \frac{d\theta_d}{d\lambda} = \frac{1}{\sqrt{1-[\lambda/\Lambda-\sin(\theta_i)]^2}}\frac{1}{\Lambda} = \frac{1}{\cos(\theta_d)} \frac{1}{\Lambda} \approx \frac{1}{\Lambda},
    \label{eq:2}
\end{equation}
with smaller $\Lambda$ resulting in larger angular dispersion. For moderate bandwidth, the dispersion will be approximately linear with wavelength. 
Any volumetric grating structure with the same index modulation and amplitude, whether made via ionization, the ponderomotive force in plasma \cite{Sheng2003plasma,Wu2005manipulating,Lee2007degenerate,Lehmann2016transient,Peng2019nonlinear}, or by heating a gas \cite{Michine2020ultra,Michel2024photochemically,Michine2024large,Michel2026entropy,Ou2026near,Singh2025holographic}, will show the same dispersive properties, although the different mechanisms have different achievable modulations and damage thresholds, and therefore utility as compression gratings.

We used 800 nm, 30 fs pulses from a 20-TW Ti:Sapphire laser (Princeton Pulsar 20 TW, Amplitude) to produce ionization gratings in two distinct configurations and measure their angular dispersion and angular bandwidth, as illustrated in Fig.~\ref{fig:1}. Figure~\ref{fig:1}(a) shows the configuration for characterizing angular dispersion. The source was split into pump and signal beams, with equal central wavelength ($\lambda_p = \lambda_0 = 800 \ \mathrm{nm}$). Since the pump and signal wavelengths were the same, the Bragg angle for the signal beam was the same as the pump crossing half-angle, $\theta_B = \theta_p = 2.2^\circ$, and we brought the signal beam out of the horizontal plane to ensure the beams could intersect without overlapping in the near field, as shown in the gas-cell side view. The resultant grating period was $10.2 \ \mathrm{\upmu m}$. The pumps and signal were focused with $f$-numbers of 40 ($F_p$) and 150 ($F_0$), respectively. The transverse size of the signal beam ($w_0$) was around $150 \ \mathrm{\upmu m}$. The transverse size of the pump beams ($w_p$), and thus the grating, was adjusted by moving the location of the pump foci longitudinally, resulting in a grating diameter of around $400 \ \mathrm{\upmu m}$. The intensity of the pump beams were about $1 \times 10^{14} \ \mathrm{W/cm^2}$ and the intensity of the signal beam was about $3 \times 10^{14} \ \mathrm{W/cm^2}$. The length of the grating was set by the gas cell to be approximately $L = 1.7 \ \mathrm{mm}$.  Carbon dioxide was used as the target gas because it has a relatively low ionization threshold, allowing large gratings with minimal pump energy. Using interferometry, we measured the plasma electron number density to be around $3.5 \times 10^{17} \ \mathrm{cm^{-3}}$. At this density, the plasma index of refraction is given by $1-n_{\textup{plasma}} \approx 1\times10^{-4}$, which supports an $n_1$ of approximately $0.5 \times 10^{-4}$.

\begin{figure}
    \centering
    \includegraphics[width=1\linewidth]{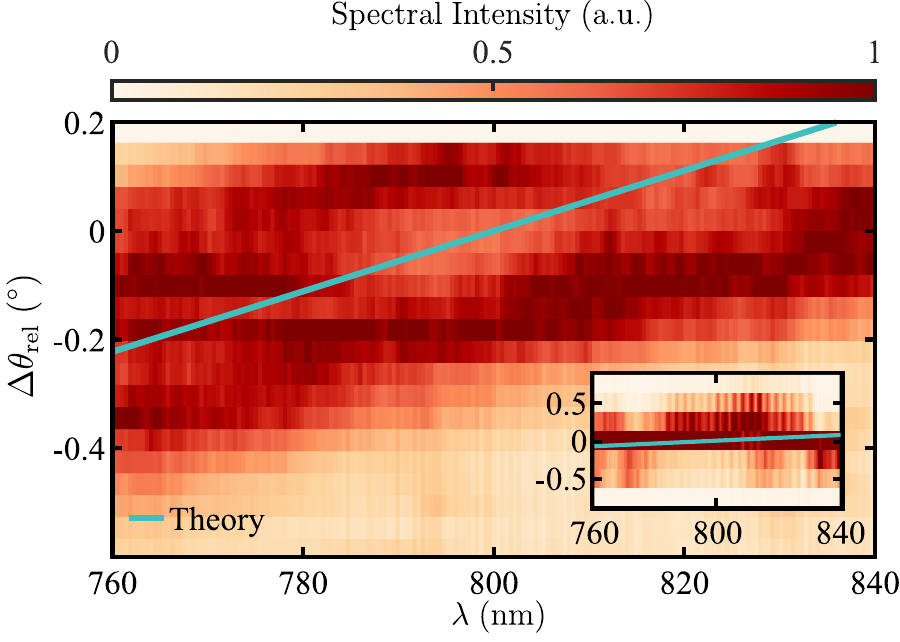}
    \caption{Experimental angular dispersion measurements for a grating with $\Lambda \approx 10.2 \ \mathrm{\upmu m}$. The teal line is calculated from the grating equation with the Bragg angle of the central wavelength used as the incident angle. The inset shows that a grating with a larger period ($\Lambda \approx 30.6 \ \mathrm{\upmu m}$) does not produce a measurable angular dispersion.}
    \label{fig:2}
\end{figure}

To measure the angular dispersion, the diffracted beam was routed to a fiber spectrometer on a translation stage. The fiber position was scanned across the diffracted beam, and an average spectrum was calculated for each position from 2000 individual spectral measurements. A corresponding diffraction angle ($\theta_d$) was calculated for each stage position. Figure \ref{fig:2} shows these average spectra for each diffraction angle $\Delta\theta_{\textup{rel}} = \theta_d - \theta_{d,0}$, where $\theta_{d,0}$ is the diffraction angle for the cental wavelength. To compensate for the spectrum of the laser, the spectral intensities in Fig.\ \ref{fig:2} were normalized by the maximum intensity at each wavelength, allowing the change in diffraction due to angular dispersion to be clearly seen. The blue line marks the predicted dispersion from the grating equation. Figure 2 shows that our experimental results are in agreement with the theory prediction of an angular dispersion of $d\theta_d/d\lambda \approx 0.005^{\circ}/\mathrm{nm}$. For comparison, the inset of Fig.\ \ref{fig:2} shows dispersion measurements for a grating with a much larger grating period ($\Lambda \approx 30.6 \ \mathrm{\upmu m}$), where we see no significant dispersion.

\begin{figure}
    \centering
    \includegraphics[width=1\linewidth]{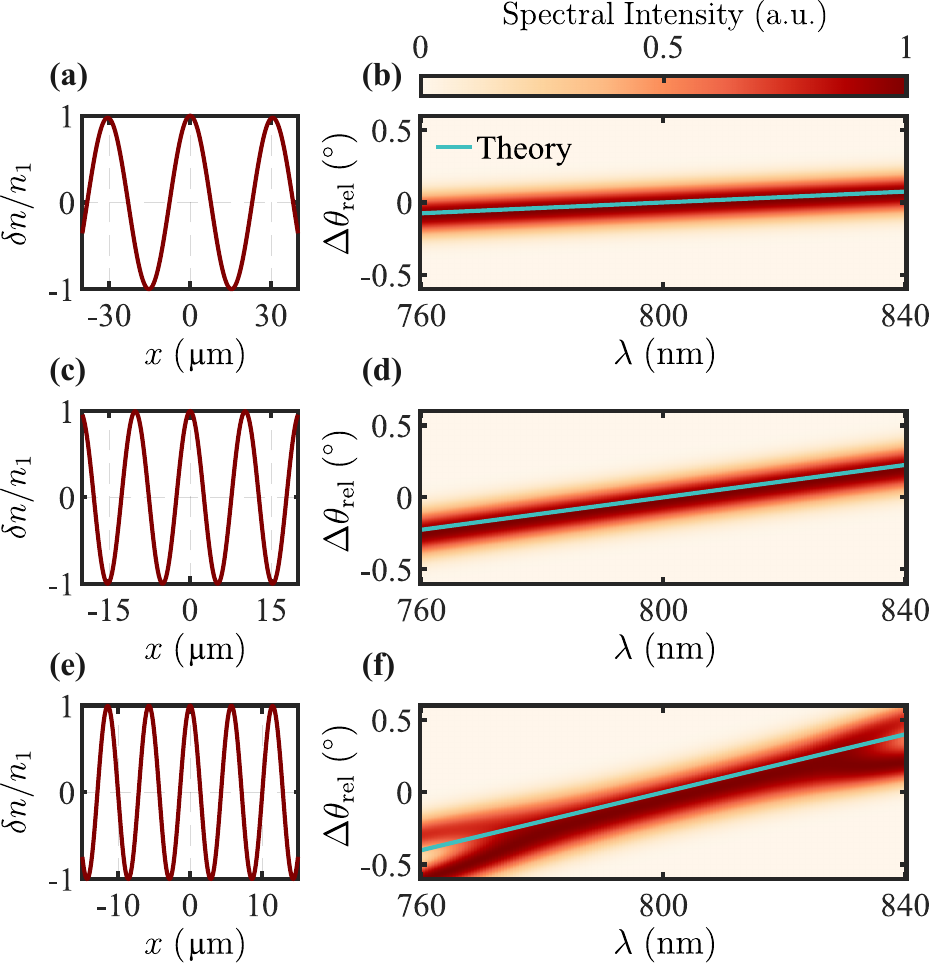}
    \caption{Simulated sinusoidal refractive index modulation profile normalized by $n_1 = 0.5 \times10^{-4}$ for gratings where (a) $\Lambda = 30.6 \ \mathrm{\upmu m}$, (c) $\Lambda = 10.2 \ \mathrm{\upmu m}$, (e) $\Lambda = 5.7 \ \mathrm{\upmu m}$. Simulated angular dispersion measurements for a Gaussian beam incident on gratings where (b) $\Lambda = 30.6 \ \mathrm{\upmu m}$, (d) $\Lambda = 10.2 \ \mathrm{\upmu m}$, (f) $\Lambda = 5.7 \ \mathrm{\upmu m}$. The teal line in each dispersion plot is calculated from the grating equation with the Bragg angle of the central wavelength used as the incident angle.}
    \label{fig:3}
\end{figure}

Additionally, we used a 2D paraxial solver to simulate pulse propagation through these gratings. We propagated a beam with a central wavelength of $\lambda_0 = 800\ \mathrm{nm}$ and a full-width at half-maximum (FWHM) spectral bandwidth of $\Delta\lambda_{\textup{FWHM}} = 0.1\lambda_0$ through gratings with three different grating periods: (A) $\Lambda = 30.6 \ \mathrm{\upmu m}$, (B) $\Lambda = 10.2 \ \mathrm{\upmu m}$ and (C) $\Lambda = 5.7 \ \mathrm{\upmu m}$. We modeled our gratings as a medium with a refractive index that could be separated into uniform and nonuniform terms, $n(x) = n_0 + \delta n(x)$, with $\delta n$ having a modulation amplitude of $n_1$. Simulations (A) and (B) used a resolution of $0.8 \ \Delta 
x/\lambda$ in the transverse direction while simulation (C) used a resolution of $0.5 \ \Delta x/\lambda$. All simulations used a resolution of $100 \ \Delta z/\lambda$ within the plasma, and all resolutions were validated for convergence. Figures \ref{fig:3}(a), \ref{fig:3}(c), and \ref{fig:3}(e)  show the index modulation distribution with $n_1 = 0.5 \times 10^{-4}$ for the three different gratings. The grating length, beam $f$-numbers, and beam diameters for the simulations were set to the experimental values: $L = 1.7 \ \mathrm{mm}$, $F_p = 40$, $F_0=150$, $w_p = 400 \ \mathrm{\upmu m}$, and $w_0 = 150 \ \mathrm{\upmu m}$. Figures \ref{fig:3}(b), \ref{fig:3}(d), and \ref{fig:3}(f) show the simulated angular dispersion when our broadband pulse was decomposed into 161 underlying frequency components, each incident at the Bragg angle of the central frequency, $\theta_i = \theta_B(\lambda = \lambda_0)$. Like the experimental angular dispersion measurement, the diffraction angles, $\theta_d$, were centered about the theoretical diffraction angle of the central frequency, $\theta_{d,0}$.  From Fig.\ \ref{fig:3}, we can see that the angular dispersion from each grating closely follows analytical predictions, increasing with decreasing grating period.

We also compare the optical performance of ionization gratings to ideal volume transmission gratings by measuring the angular bandwidth and the variation of the diffraction angle with angle of incidence. The angular bandwidth can be found by generalizing Eq.~\ref{eq:0} \cite{Yeh1993introduction}:
\begin{equation}
    \eta(\theta_i,\lambda) = \frac{\sin^2\left(\kappa L\sqrt{1+(\Delta\alpha/2\kappa)^2}\right)}{1+(\Delta\alpha/2\kappa)^2},
    \label{eq:3}
\end{equation}
where $\Delta\alpha = -2\pi\left(\theta_i-\theta_B\right)/\Lambda$ is the phase mismatch resulting from the difference between the incident angle and the Bragg angle for wavelength $\lambda$. Equation (\ref{eq:3}) leads to an expression for the angular bandwidth of a volume transmission grating \cite{Yeh1993introduction,Edwards2022plasma}:
\begin{equation}
    \Delta\theta_{g,\textup{FWHM}} \approx \frac{\Lambda}{L_{\textup{opt}}} = \frac{n_1}{\sin\left(\theta_B\right)\cos\left(\theta_B\right)},
    \label{eq:4}
\end{equation}
where $\Delta\theta_{g,\textup{FWHM}}$ is the FWHM bandwidth and $L_{\textup{opt}} = \pi/(2\kappa)$ is the length of the grating that results in maximized efficiency when $\Delta\alpha = 0$. The spectral bandwidth can also be derived from Eq.~\ref{eq:3}. Rewriting the phase mismatch in terms of a wavelength difference rather than an angle difference gives $\Delta\alpha = -\pi(\lambda-\lambda_0)/(\Lambda^2n_0\cos\left[\theta_B\right])$ \cite{Yeh1993introduction,Edwards2022plasma} and with $n_0 \approx 1$:
\begin{equation}
    \frac{\Delta\lambda_{g,\textup{FWHM}}}{\lambda_0} \approx \frac{4n_1\Lambda^2}{\lambda_0^2} = \frac{n_1}{\sin^2\left(\theta_B\right)},
    \label{eq:5}
\end{equation}
where $\Delta\lambda_{g,\textup{FWHM}}$ is the FWHM spectral bandwidth. 
Both the angular bandwidth and the spectral bandwidth are directly proportional to the grating period and can be increased by increasing $n_1$, allowing smaller $L$ for the same diffraction efficiency. It is important that the spectral bandwidth of the grating be greater than or equal to the bandwidth of the laser pulse being compressed as any narrowing of the pulse spectrum will result in inefficient compression and longer pulse duration.

\begin{figure}[b]
    \centering
    \includegraphics[width=1\linewidth]{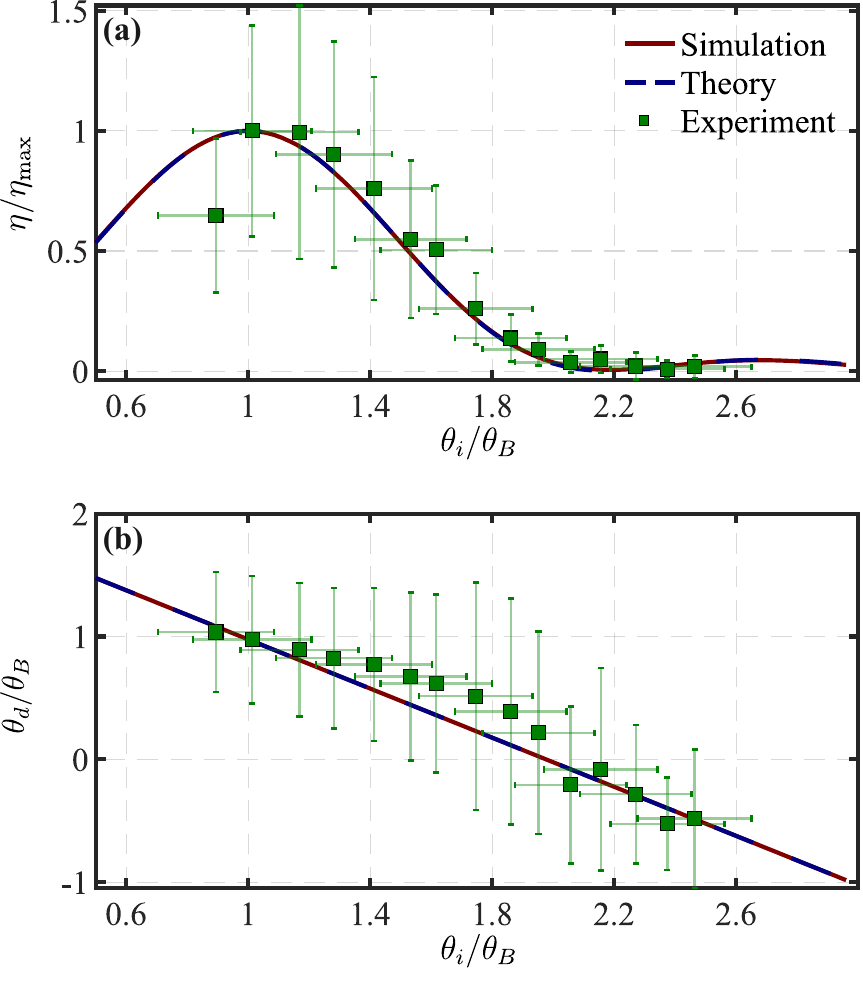}
    \caption{(a) Simulated, theoretical, and experimental diffraction efficiency versus incident angle. The experimental data points represent an average over 200 individual shots and the vertical error bars represent $\pm 1$ standard deviation. The horizontal errors bars represent the error in the measurements of the incident angle. The theoretical curve is calculated using Eq.\ \ref{eq:3}. (b) Simulated, theoretical, and experimental first-order diffraction angle versus incident angle. The horizontal and vertical errors bars represent the error in the measurements of the incident and diffraction angles, respectively. The theoretical curve is calculated using Eq.\ \ref{eq:1}.}
    \label{fig:4}
\end{figure}

We used the experimental configuration shown in Fig.\ \ref{fig:1}(b) to measure the angular bandwidth of our ionization gratings. A Teflon screen was placed after the gas cell to image both the zeroth-order and first-order diffracted signal beam. The pump beams were frequency-doubled to $\lambda_p = 400 \ \mathrm{nm}$ and crossed at $\theta_p = 0.375^{\circ}$, giving $\Lambda = 30.6 \ \mathrm{\upmu m}$. The grating length was set by the gas cell to $L = 2 \ \mathrm{mm}$. We varied the signal angle of incidence by selecting a 2-mm-diameter portion of the signal beam with an iris, then scanning the iris transversely across the beam with a translation stage. 

Simulations were also conducted with the same paraxial solver used for our angular dispersion simulations. We varied the angle of incidence from $0.5\theta_B$ to $3\theta_B$ in steps of $0.025\theta_B$ for a broadband pulse with central wavelength $\lambda_0 = 800 \ \mathrm{nm}$ and FWHM bandwidth $\Delta\lambda_{\textup{FWHM}} = 0.1\lambda_0$ that was decomposed into 183 frequency components. $1.1 \ \mathrm{to} \ 3.3 \ \Delta x/\lambda$ resolution was used in the transverse direction and $100 \ \Delta z/\lambda$ resolution was used within the plasma. The beam properties and grating geometry were equivalent to those used in the angular bandwidth experiment: $F_p = 100$, $F_0 = 500$, $w_p = 400 \ \mathrm{\upmu m}$, $w_0 = 275 \ \mathrm{\upmu m}$, $\lambda_p = 400 \ \mathrm{nm}$, $L = 2 \ \mathrm{mm}$, and $2\theta_p = 0.75^{\circ}$. The diffraction angle for each frequency was measured by using the transverse component ($k_{d,x}$) of the diffracted beam k-vector ($k_d$), $\theta_d = \arcsin(k_{d,x}/k_d)$, while the diffraction efficiency was measured by integrating the intensity of the first-order diffracted beam and dividing it by the total incident power. 

\begin{figure}
    \centering
    \includegraphics[width=1\linewidth]{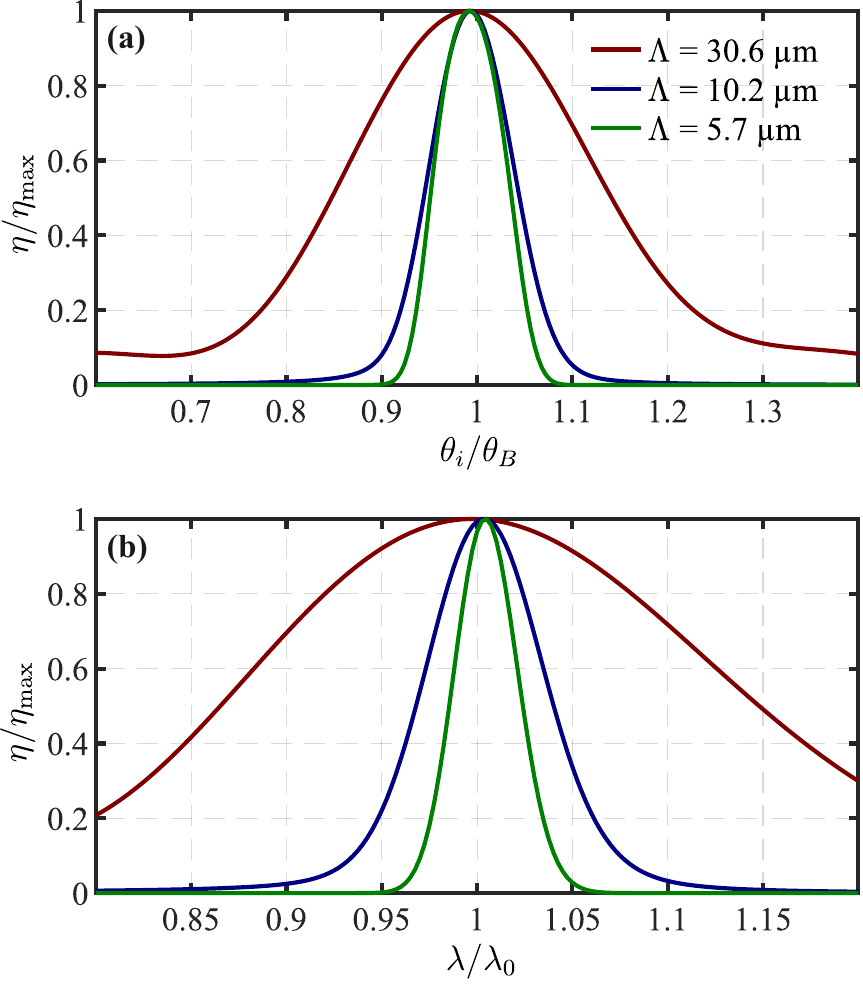}
    \caption{Simulated normalized diffraction efficiency as a function of (a) incident angle and (b) wavelength for gratings with grating periods of $\Lambda = 30.6 \ \mathrm{\upmu m}$, $\Lambda = 10.2 \ \mathrm{\upmu m}$, and $\Lambda = 5.7 \ \mathrm{\upmu m}$.}
    \label{fig:5}
\end{figure}

Figure \ref{fig:4}(a) shows the simulated, theoretical, and experimental diffraction efficiencies as a function of the signal incident angle. Each diffraction efficiency curve was normalized by its respective maximum diffraction efficiency, $\eta/\eta_{\textup{max}}$. From Fig.~\ref{fig:4}(a), we can see that the simulated, theoretical, and experimental results peak near the Bragg angle of the grating. As the angle deviates from the Bragg angle, the efficiency drops. Figure\ \ref{fig:4}(a) shows the diffraction efficiency falling to $0.5\eta_{\textup{max}}$ at $\theta_i = 1.6\theta_B$; however, this does not represent the grating’s angular bandwidth. The finite beam divergence and bandwidth of the signal beam in addition to a non-optimal grating length introduce phase mismatches that reduce the peak efficiency and broaden the angular response. These effects are included in both the simulations and theory, explaining the strong agreement among all the curves. Using the results from simulations and experiment, we obtain an index modulation of $n_1 = 0.125 \times 10^{-4}$, which, when substituted into Eq.~\ref{eq:4}, gives an angular bandwidth of $\Delta\theta_{g,\textup{FWHM}} = 0.07\theta_B$. Figure \ref{fig:4}(b) shows the experimental first-order diffraction angle as a function of the signal incident angle. The simulated and theoretical diffraction angles were also plotted for comparison. We can see that the diffractive behavior of our gratings is well-predicted by theory and simulations.

To explore the dependence of bandwidth on grating period, we simulated propagation through three gratings with different periods. Each grating had $n_1 = 0.5 \times 10^{-4}$ and $L = 8 \ \mathrm{mm} \approx L_{\textup{opt}}$. The angular bandwidth simulations modeled a broadband pulse with a Gaussian spectrum of width $\Delta\lambda_{\textup{FWHM}} = 0.1\lambda_0$ while the spectral bandwidth simulations used a spectrum of width $\Delta\lambda_{\textup{FWHM}} = 0.4\lambda_0$. The spectrum was decomposed into 81 frequency components for the angular bandwidth simulations and 161 frequency components for the spectral bandwidth simulations. The simulations used a resolution of 2.2 to 2.6, 1.3 to 1.6, and 0.9 to 1.1 $\Delta x/\lambda$ in the transverse direction for gratings with periods of $30.6$, $10.2$, and $5.7 \ \mathrm{\upmu m}$, respectively. All simulations used $100 \ \Delta z/\lambda$ resolution within the plasma. Figure \ref{fig:5}(a) shows the simulated efficiencies as a function of the signal incident angle while Fig.\ \ref{fig:5}(b) shows the efficiencies as a function of the signal incident wavelength. The grating with the largest grating period has over twice the angular and spectral bandwidth of the gratings with smaller grating period. Additionally, as the grating period gets smaller, the bandwidth decreases as expected.

\begin{figure} [t]
    \centering
    \includegraphics[width=1\linewidth]{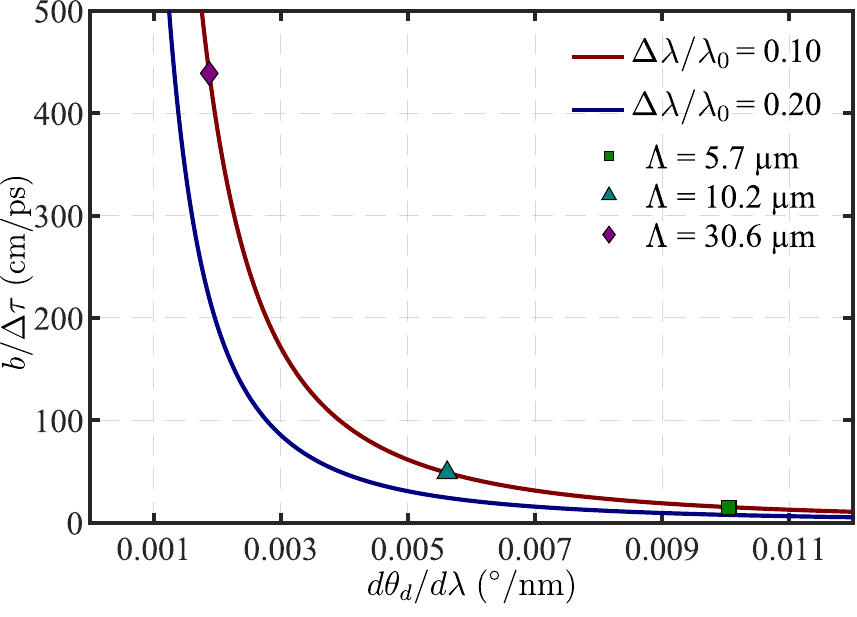}
    \caption{Separation distance per unit time of compression as a function of angular dispersion for beams with two different spectral bandwidths. The green, teal, and purple data points represent the amount of $b/\Delta\tau$ needed for compressors built using $5.7$, $10.2$, and $30.6 \ \mathrm{\upmu m}$, respectively.}
    \label{fig:6}
\end{figure}

Both the angular dispersion and grating bandwidths depend on $\Lambda$, resulting in a tradeoff where large bandwidths favor large grating periods while large angular dispersion favors small grating periods. Balancing these effects is important for building a compact, plasma-based laser compressor. The compression of a pulse with bandwidth $\Delta\lambda $ by an amount $\Delta \tau$ can be calculated using \cite{Yang1985femtosecond,Edwards2022plasma}:
\begin{equation}
    {\Delta\tau} = \frac{b\lambda_0}{c}\frac{\Delta\lambda}{\Lambda^2}\frac{1}{1-[\lambda/\Lambda-\sin(\theta_i)]^2},
    \label{eq:6}
\end{equation}
where $b$ is the separation distance between two gratings and $c$ is the speed of light. Equation (\ref{eq:6}) can be rewritten to calculate the amount of separation needed per unit time to fully compress the pulse,
\begin{equation}
    \frac{b}{\Delta\tau} = \frac{c}{\lambda_0^2}\left(\frac{\Delta\lambda}{\lambda_0}\right)^{-1}\left(\frac{d\theta_d}{d\lambda}\right)^{-2}.
    \label{eq:7}
\end{equation}
The equation above shows that achieving high angular dispersion can significantly reduce a compressor's size, which is why we aim to design efficient gratings with maximum possible dispersion. 

Figure \ref{fig:6} plots the separation distance per unit time of compression as a function of the angular dispersion for a beam with central wavelength $\lambda_0 = 800 \ \mathrm{nm}$ and two different bandwidths. We calculated the theoretical dispersion of gratings with periods of $30.6 \ \mathrm{\upmu m}$, $10.2 \ \mathrm{\upmu m}$, and $5.7 \ \mathrm{\upmu m}$  using Eq.\ (\ref{eq:2}) and plotted the points on the line corresponding to the FWHM bandwidth of our laser ($\Delta\lambda_{\textup{FWHM}} = 0.1\lambda_0$) to determine how much separation distance would be needed to build a compressor. To compress a $10 \ \mathrm{ps}$ pulse down to its Fourier-transform limit, a compressor built with gratings with a $10.2 \ \mathrm{\upmu m}$ period like those generated in our angular dispersion experiments would approximately require $490 \ \mathrm{cm}$ of separation between the two gratings. Compressors built using gratings with periods of $30.6 \ \mathrm{\upmu m}$ and $5.7 \ \mathrm{\upmu m}$  would require $4400 \ \mathrm{cm}$ and $150 \ \mathrm{cm}$, respectively. As seen in Fig.\ \ref{fig:5}(b), though, the spectral bandwidth of a grating with a period of $5.7 \ \mathrm{\upmu m}$ is too small to efficiently diffract the entire FWHM bandwidth of a pulse with $\Delta\lambda_{\textup{FWHM}} = 0.1\lambda_0$, resulting in inefficient compression. If a beam with a broader bandwidth of $\Delta\lambda_{\textup{FWHM}} = 0.2\lambda_0$ was incident on gratings with $30.6$, $10.2$, and $5.7 \ \mathrm{\upmu m}$ periods, then the separation distance would be $2200 \ \mathrm{cm}$, $240 \ \mathrm{cm}$, and $75 \ \mathrm{cm}$, respectively. However, gratings with higher index modulation amplitudes, and thus higher grating bandwidths, would have to be used to efficiently diffract the entire bandwidth and compress the pulse fully.

In summary, we have experimentally quantified the angular dispersion and bandwidth properties of different ionization gratings and compared our findings to theoretical and computational predictions. Our experimental results demonstrate that plasma gratings have the expected dispersive properties needed for future plasma-based pulse compression schemes. Given their larger damage thresholds, these gratings present a viable pathway towards the development of compact laser compressors that could enable future hundreds-of-petawatts-to-exawatt-class laser facilities. These results provide a foundation for plasma-based diffractive optics for the next generation of high-power lasers.

This work was supported by the U.S. Department of Energy under Grants DE-SC0025497, and DE-SC002557, the U.S. National Science Foundation under Grants PHY-2206711, PHY-2308641, and PHY-2541940, the National Nuclear Security Administration under Grant DE-NA0004130, and the Gordon and Betty Moore Foundation under Grant DOI 10.37807/gbmf12255, by the National Science Foundation Graduate Research Fellowship under Grant No. DGE-2146755 (V. M. P-R.) and Grant No. DGE-2444107 (I. T-G.). Lawrence Livermore National Laboratory is operated by Lawrence Livermore National Security, LLC, for the U.S. Department of Energy, National Nuclear Security Administration under Contract No. DE-AC52-07NA27344.

\bibliography{references}

\end{document}